\documentclass[conference]{IEEEtran}
\IEEEoverridecommandlockouts
\usepackage{cite}
\usepackage{amsmath,amssymb,amsfonts}
\usepackage{algorithmic}
\usepackage{graphicx}
\usepackage{textcomp}
\usepackage{subcaption}
\usepackage{xcolor}
\usepackage{makecell}
\usepackage{multirow}
\usepackage{array}

\def\BibTeX{{\rm B\kern-.05em{\sc i\kern-.025em b}\kern-.08em
    T\kern-.1667em\lower.7ex\hbox{E}\kern-.125emX}}
\begin{document}

\title{Explicit-NeRF-QA: A Quality Assessment Database for Explicit NeRF Model Compression}

\author{Anonymous Author(s)}


 \author{
     Yuke Xing\textsuperscript{\#1},
     Qi Yang\textsuperscript{*2},
     Kaifa Yang\textsuperscript{\#1},
     Yilin Xu\textsuperscript{\#1},
     Zhu Li\textsuperscript{*3} \\
     \# Cooperative MediaNet Innovation Center, Shanghai Jiao Tong University, Shanghai, China \\
     * Department of Computer Science and Electrical Engineering, University of Missouri-Kansas City, USA \\
     \textsuperscript{1}\{xingyuke-v, sekiroyyy, yl.xu\}@sjtu.edu.cn \\
     \textsuperscript{2}littlleempty@gmail.com \\
     \textsuperscript{3}lizhu@umkc.edu \\
     \vspace{-2em}
 }

\maketitle

\begin{abstract}
In recent years, Neural Radiance Fields (NeRF) have demonstrated significant advantages in representing and synthesizing 3D scenes. Explicit NeRF models facilitate the practical NeRF applications with faster rendering speed, and also attract considerable attention in NeRF compression due to its huge storage cost. To address the challenge of the NeRF compression study, in this paper, we construct a new dataset, called Explicit-NeRF-QA. We use 22 3D objects with diverse geometries, textures, and material complexities to train four typical explicit NeRF models across five parameter levels. Lossy compression is introduced during the model generation, pivoting the selection of key parameters such as hash table size for InstantNGP and voxel grid resolution for Plenoxels. By rendering NeRF samples to processed video sequences (PVS), a large scale subjective experiment with lab environment is conducted to collect subjective scores from 21 viewers. The diversity of content, accuracy of mean opinion scores (MOS), and characteristics of NeRF distortion are comprehensively presented, establishing the heterogeneity of the proposed dataset. The state-of-the-art objective metrics are tested in the new dataset. Best Person correlation, which is around 0.85, is collected from the full-reference objective metric. All tested no-reference metrics report very poor results with 0.4 to 0.6 correlations, demonstrating the need for further development of more robust no-reference metrics. The dataset, including NeRF samples, source 3D objects, multiview images for NeRF generation, PVSs, MOS, is made publicly available at the following location: 
https://github.com/YukeXing/Explicit-NeRF-QA.

\end{abstract}


\section{Introduction}
Neural Radiance Fields (NeRF)~\cite{nerf} is a method to reconstruct 3D scenes from 2D sparse multiview images, which has an impressive ability to synthesize novel views with high realism. NeRF variants could be categorized into two types: implicit and explicit. Implicit NeRF methods encode 3D scenes implicitly by sampling a large number of spatial points in the rendering process based on MLP networks, resulting in high computational cost and slow convergence speed. In contrast, explicit NeRF methods divide 3D scenes into voxel grids and store the local features in each grid cell, which significantly reduces the computational load during the training and rendering process, thus greatly promoting the development of NeRF practical applications. 


One disadvantage of explicit NeRF models is that they inevitably increase the memory and storage overhead. Therefore, researchers have proposed multifarious strategies to reduce the size of explicit NeRF methods pivoting to a series of compression methods~\cite{NeRFcomp1,NeRFcomp2,NeRFcomp3,NeRFcomp4,NeRFcomp5}. Similarly to image/video compression, NeRF compression also incurs distortion and impacts perceptual quality. To find the most suitable compression strategies, effective NeRF quality assessment (NeRF-QA) metrics are needed. However, because of the lack of research on NeRF-QA, most researches still use traditional image quality metrics to assess NeRF models' quality, which may neglect unique distortions of the NeRF models and are resulting in inaccurate prediction. Therefore, a new dataset with diverse types of NeRF distortion samples and substantial scale is firstly needed for designing and evaluating NeRF-QA metrics. An effective metric can facilitate the study of NeRF compression, revealing that a NeRF-QA dataset is consequently necessary for the study of NeRF compression.

As we know, there have been three studies on NeRF-QA dataset. \cite{NeRF-QA-1} collected 14 real scenes, together with the scenes in LLFF~\cite{LLFF}, they rendered NeRF models into videos and performed subjective experiments. \cite{NeRF-QA-2} used 4 real scenes from ``Tanks and Temples"~\cite{Tanksandtemples} and 4 synthetic scenes from NeRF\_Synthetic~\cite{nerf}. For each kind of scene, they chose 4 NeRF methods to synthesize videos and obtain perception quality scores. \cite{NeRF-QA-3} considered 8 real scenes and 8 synthetic scenes on several NeRF methods. The above works are the pioneer of NeRF-QA and contribute significantly to the NeRF study. However, the above datasets have the disadvantages of small-scale or scanty distortion types with only one quality level for each NeRF model. Considering the rapid development of the NeRF study, it is necessary to construct a new dataset that provides diverse content, rich distortions, and trustworthy mean opinion scores (MOS).


In this paper, we create a new dataset called Explicit-NeRF-QA. We use 22 synthetic scenes as reference to generate NeRF models. All NeRF methods belong to mainstream and widely studied explicit models, i.e., InstantNGP \cite{Instantngp}, DVGO \cite{DVGO}, Plenoxels \cite{Plenoxels}, and TensoRF \cite{TensoRF}. Each NeRF model has five quality levels corresponding to different compression levels by controlling key model parameters, such as the hash table size for InstantNGP. In all, there are $22\times4\times5 = 440$ samples in Explicit-NeRF-QA. Both reference and distorted NeRF samples are rendered into processed video sequences (PVS) to conduct subjective experiments. We split the dataset into two parts: 20 samples used in training sessions and 420 samples in rating sessions. The double stimulus impairment scale is used to collect subjective scores. After removing outliers, we analyze the content diversity and MOS accuracy, and illustrate that NeRF models have distortions distinct from traditional 2D images, such as fogging, ripple, and wave, as defined in Section~\ref{sec:distortion}. In addition, NeRF models have different response to objects' materials, specular reflective material is the most challenging type for NeRF generation and lossy compression. We test 15 state-of-the-art (SOTA) objective metrics. The highest Person and Speraman correlations are around 0.85, which are reported from full-reference metrics, while all the tested no-reference metrics show very low correlations, indicating the need for further development of more robust no-reference metrics for NeRF models. 

\section{Dataset Construction}
This section presents the selection of source content, the generation of multiview images, the selection and training of NeRF models, and the subjective experiments.

\subsection{Source Content Selection}
\begin{figure}[htbp]
\vspace{-1em}
    \centering 
    \includegraphics[width=1\columnwidth]{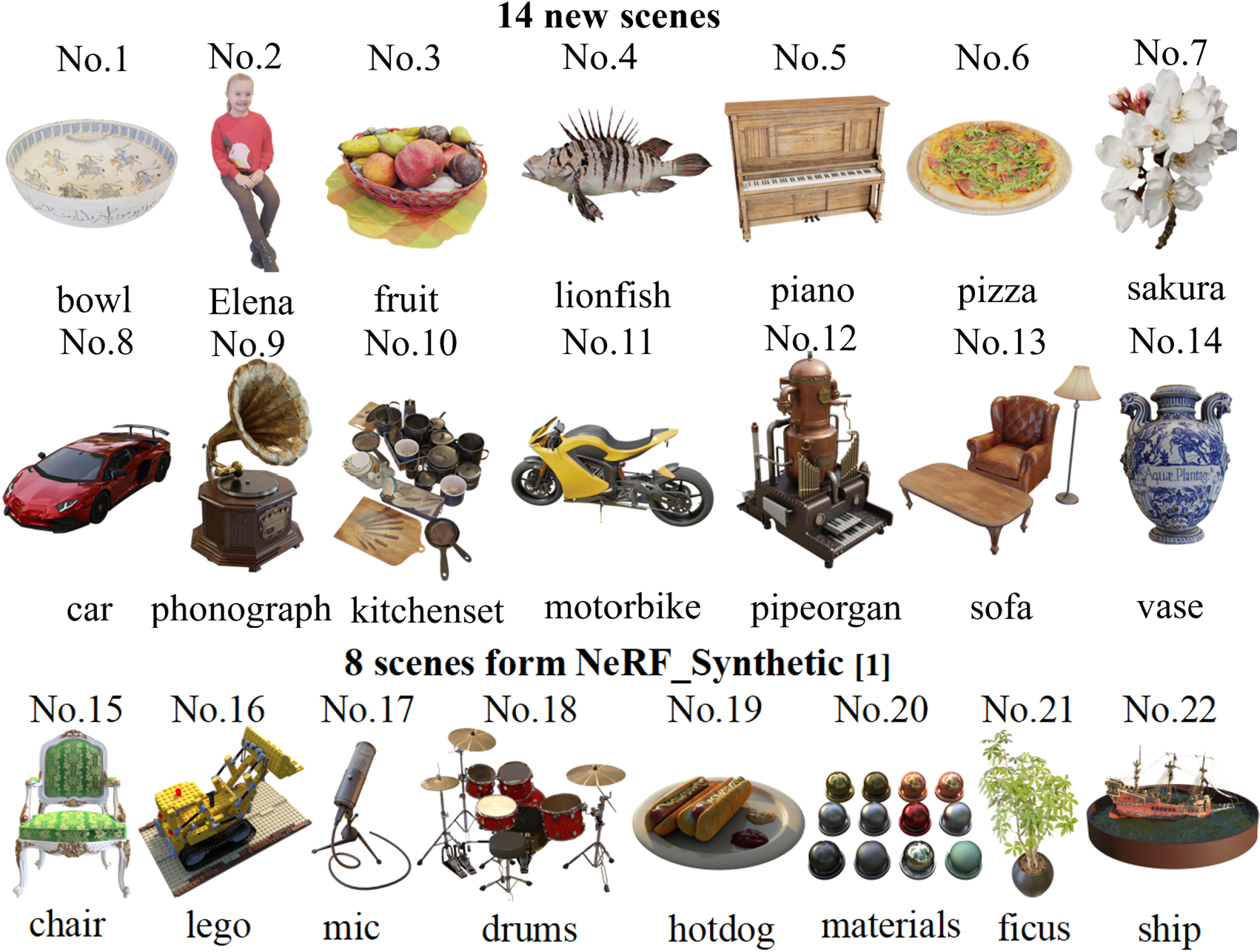} 
    \caption{The source content in Explicit-NeRF-QA}
    \label{fig:source}
    \vspace{-1em}
\end{figure}

The performance of NeRF models is correlated with material categories, therefore, we additionally consider material diversity in source content selection. Specifically, we use 22 different contents as source, of which eight from previous work \cite{nerf} and 14 new high-quality 3D scenes carefully selected from SketchFab\cite{sketchfab} with diverse categories, materials, geometry and texture complexity. 
The 22 selected scenes are shown in Fig.~\ref{fig:source}.

For the 14 new high-quality 3D scenes, we categorize their materials into three types based on the properties of light reflection: \textbf{1) Specular reflective materials (SRM)}: Objects possess highly smooth surfaces that can produce clear reflections. The reflected light highly depends on the position of observer and light source. \textbf{2) Diffuse reflective materials (DRM)}: Objects feature rough surfaces that scattering incident light uniformly in multiple directions. The color and brightness remain consistent viewed from any angle. \textbf{3) Glossy reflective materials (GRM)}: Objects have common surfaces (between smooth and rough) that can produce blurred reflections, the clarity depends on the surface's roughness.
Among the 14 additinoal 3D scenes in Fig.~\ref{fig:source}, No.1-7 belong to DRM, No.8 and 9 belong to SRM, and No.10-14 belong to GRM.
\subsection{Multiview Image Generation}
The generated multiview images consist of three parts based on the requirements and rules of training NeRF models: training, testing and validation set. The training set is used as input for training NeRF models, testing set is used to calculate objective metrics, and validation set is used to render videos for subjective experiments. For training and testing sets, we randomly select 100 viewpoints on the sphere or hemisphere around objects. Training and testing sets do not have repeated viewpoints. For validation set, we select 300 viewpoints located spirally at the upper hemisphere circling the object with uniform angular intervals. Each set includes multiview images and the corresponding camera parameter file.

\subsection{NeRF Model Selection and Training}
\subsubsection{NeRF Model Selection}
Four prevalent and representative explicit NeRF models are selected as anchor methods to generate NeRF samples:


\textbf{InstantNGP}~\cite{Instantngp}: InstantNGP employs a multi-level hash encoding structure to map multi-resolution voxel grids into different hash tables, which can realize storing learnable feature vectors more compactly.

\textbf{DVGO}~\cite{DVGO}: DVGO proposes two initialization algorithms to prevent the scene geometry from falling into local optima, and introduces a post-activation voxel grid interpolation method that generates more clear boundaries at lower grid resolutions. 

\textbf{Plenoxels}~\cite{Plenoxels} : With each grid storing volume density and spherical harmonics coefficients, Plenoxels directly optimizes these parameters through gradient backpropagation and regularization methods without relying on any neural network. 

\textbf{TensoRF}~\cite{TensoRF}: TensoRF views the voxel grid as a 4D tensor, and introduces a Vector-Matrix decomposition method to factorize the 4D tensor into multiple and compact low-rank tensor components, thus reducing storage space.

\subsubsection{NeRF Model Training}
To build a large-scale NeRF-QA dataset that includes various distortion types and quality levels, we propose five compression parameter levels (PL) for each NeRF method to produce samples with hierarchical model size and quality. The key parameters selected for each NeRF model are shown in Table~\ref{tab:nerf_parameters}.

\begin{table}[htbp]
\vspace{-1em}
    \centering
    \caption{PL of NeRF models. L01-L05 correspond to the five PLs. The original parameter setting recommended by each model is used in L05.}
    \label{tab:nerf_parameters}
    \resizebox{\linewidth}{!}{
    \begin{tabular}{|c|c|c|c|c|}
    \hline
    Model & Instantngp & Plenoxels & DVGO & TensoRF\\ \hline
     \makecell{Key \\ Parameters}& \makecell{Hash Table \\ Size $T$} & \makecell{Voxel Grid \\ Resolution} & \makecell{Voxel Grid \\Resolution\\  (coarse, fine)} & \makecell{Voxel Grid \\Resolution,\\Component Count} \\ \hline
    L01 & $2^6$ & $32^3$ & $[10^3, 10^3]$ & $[15^3, \text{Comp:}48]$ \\ \hline
    L02 & $2^8$ & $64^3$ & $[10^3, 32^3]$ & $[25^3, \text{Comp:}48]$ \\ \hline
    L03 & $2^{10}$ & $128^3$ & $[15^3, 32^3]$ & $[40^3, \text{Comp:}96]$ \\ \hline
    L04 & $2^{13}$ & $256^3$ & $[32^3, 64^3]$ & $[100^3, \text{Comp:}96]$ \\ \hline
    L05 & $2^{19}$ & $512^3$ & $[100^3, 160^3]$ & $[300^3, \text{Comp:}192]$ \\ \hline
    \end{tabular}}
    \vspace{-2em}
\end{table}

\subsection{PVS Generation}
To conduct subjective experiments, each NeRF sample is processed into PVS. By employing predefined camera trajectories in the validation set, 300 frames are rendered that circle the object with uniform angular intervals at a resolution of $800\times 800$. Subsequently, with a frame rate of 30 and constant rate factor of 10 to guarantee visually lossless encoding~\cite{PVS}, these images are compiled into PVS through FFMPEG utilizing libx265. Each PVS has a duration of 10 seconds.

\subsection{Subjective Experiment}
\subsubsection{Training and Evaluation}
To guarantee the reliability of the collected subjective scores, we employ the "lionfish" (No.11 in Fig.~\ref{fig:source}) to establish a training session, following the same method proposed in~\cite{train}. In the rating session, an 11-level impairment scale proposed by ITU-TP.910~\cite{ITU1(siti)} is used as the rating method. A double stimulus impairment scale method is utilized, with reference and NeRF synthesized PVSs displayed side-by-side. The experiment was carried out using a 27-inch AOC Q2790PQ monitor in an indoor laboratory environment under standard lighting conditions. To align with the PVSs' format, the monitor's display resolution is set to $1600\times 800$. To prevent visual fatigue caused by too long experiment time, 420 PVSs are randomly divided into 6 smaller groups.

\subsubsection{Outlier Removal}
To filter outliers from the raw subjective scores, two consecutive steps are used. First, we add an extremely low-quality PVS and a duplicated PVS to each rating session, known as trapping samples. After collecting subjective scores, outliers are identified and excluded based on the ratings to these trapping samples. Subsequently, ITU-R BT.500~\cite{ITU2} is applied for the secondary detection and elimination of outliers. Finally, two outliers are identified and removed from the initial subjective scores.
\section{Dataset Validation}

\subsection{Diversity of Content}
\begin{figure}[htbp]
\vspace{-1em}
    \centering 
    \includegraphics[width=0.95\columnwidth]{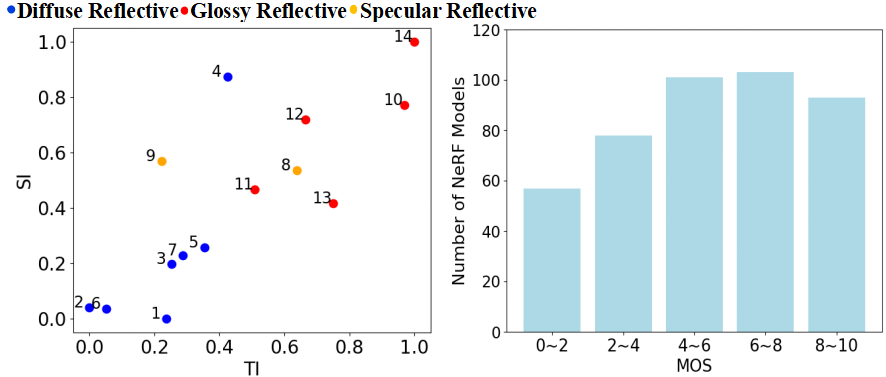} %
    \caption{Left: SI vs. TI . Right: MOS Distribution}
    \label{fig:MOS}
    \vspace{-1.5em}
\end{figure}

To validate the diversity of the source content, we measure spatial perceptual information (SI), temporal perceptual information (TI)~\cite{ITU1(siti)}, and types of materials of the 14 newly selected scenes. Reference PVSs are used to calculate SI and TI. The results are shown in Fig.~\ref{fig:MOS}. The relatively uniform distribution of the scatter points indicates the diversity of the source content.


\subsection{Analysis of MOS}

To verify whether the proposed database covers a wide range of quality levels, the MOS distribution is presented in Fig.~\ref{fig:MOS}, where for each score segment, Explicit-NeRF-QA has at least 50 samples, showcasing the MOSs are evenly distributed in the whole quality range. 

\begin{figure}[htbp]
\vspace{-1em}
    \centering 
    \includegraphics[width=0.95\columnwidth]{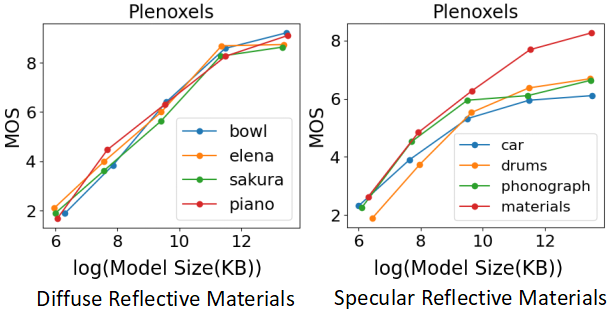} %
    \caption{MOS $vs$ log(model size(KB))}
    \label{fig:size}
    \vspace{-2em}
\end{figure}
To verify the accuracy of MOS, we examine the correlation between MOS and model size (KB). We use results of Plenoxels as examples, and pick four objects belonging to DRM (i.e.,bowl, elena, sakura, piano) and SRM (i.e., car, drums, phonograph, materials) separately. The results are shown in Fig.~\ref{fig:size}. For better observation, we apply the common logarithm transformation to the model size. Considering a larger NeRF model is supposed to indicate more detailed visual features and thus higher MOS, the results report that the curves present perfect monotonicity, validating the accuracy of the MOS scores and the reasonable settings of the PLs. For DRM, the MOS grow gradually and eventually reach 8-10 scores. While for SRM, most samples' MOS do not have notable growth with the increasing model size after reaching MOS around 5 to 7. Even under the optimal parameter setting L05, they still cannot achieve higher performance.
The result shows that human observers are highly sensitive to the distortion produced by NeRF on smooth surfaces, and current NeRF models are still not perfect and under study, which may unavoidably introduce artifact when the situation is challenging, such as synthesizing 3D scenes containing SRM.

\section{NeRF Distortion Analysis}
\label{sec:distortion}
By observing 440 NeRF PVSs and gathering feedback from participants in subjective experiments, we found that NeRF has some distortion types that are not available in other media forms, such as point clouds \cite{yangtmm} and meshes \cite{cui2024sjtu}, which is one of the main challenges in conducting objective quality assessments based on NeRF. We comprehensively and systematically summarize NeRF-specific artifact into 9 unique types:

\begin{figure}[htbp]
    \centering 
    \includegraphics[width=0.9\columnwidth]{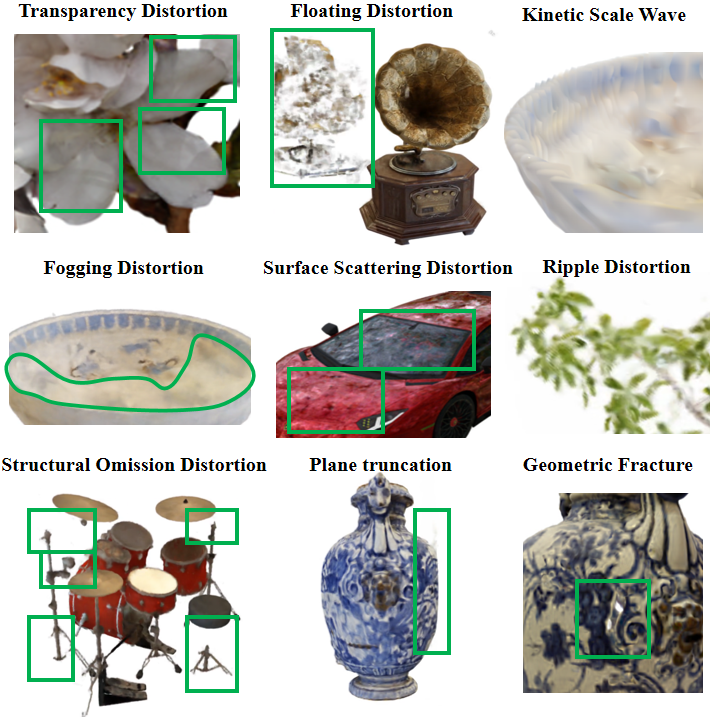} %
    \caption{Examples for NeRF Distortions}
    \label{fig:my_label}
    \vspace{-2em}
\end{figure}

\textbf{(1) Transparency Distortion}: Certain areas of an object appear more transparent than the reference one, with the content behind them partially visible.

\textbf{(2) Floating Distortion}: Typically appearing as discrete and blurry spots or blocks causing an unnatural visual break from their background.

\textbf{(3) Kinetic Scale Wave Distortion}: A scaly and repetitive texture appearance when static, and a wavy effect when in motion, giving viewers a sense of blurriness and dizziness, significantly reducing perceived quality.

\textbf{(4) Fogging Distortion}: Characterized by the appearance of fog-like or smoky, opaque, or semi-transparent coverings on the surface or nearby space of the rendered objects, obscuring the details underneath and decrease surface details and textures visual clarity.

\textbf{(5) Surface Scattering Distortion}: Often appears on the surfaces of specular reflective materials, which exhibits textures that look like cotton or cloud-like patterns, contradicting the object's actual smooth or uniform surface texture. 

\textbf{(6) Ripple Distortion}: Unnatural ripples or texture distortions make the texture of the objects look wavy or as if they have irregular creases.

\textbf{(7) Structural Omission Distortion}: Characterized by missing macro structures that disrupt the realism and continuity of the entire scene.

\textbf{(8) Plane Truncation Distortion}: Characterized by abrupt cut-offs and incomplete scenes, as if objects have been sliced or truncated along certain planes, it creates a disruptive visual experience for the viewer.

\textbf{(9) Geometric Fracture Distortion}: Appearance of geometric discontinuities and damaged effects, where the structural integrity of the object's surface is broken, disrupting the continuity of shape and texture.

\section{Objective Metrics Testing}
The performance of SOTA image/video quality metrics on PVSs are tested in this section. We select 15 objective metrics covering various types, specifically: i) full-reference image metrics, including classic metrics, PSNR, SSIM~\cite{PSNRSSIM}, MS-SSIM~\cite{MS-SSIM}, IW-SSIM~\cite{IW-SSIM}, VIF~\cite{VIF} and FSIM~\cite{FSIM}, deep-learning based metrics, inclduing LPIPS (VGG and AlexNet versions)~\cite{LPIPS} and DISTS ~\cite{DISTS}; ii) no-reference image metrics, including BRISQUE ~\cite{BRISQUE}, CLIP\_IQA~\cite{CLIP-IQA} and NIQE~\cite{NIQE}; iii) video metrics, including FovVideoVDP~\cite{FovVideoVD}, HDR-VQM~\cite{HDR-VQM} and VMAF~\cite{VMAF}.

\subsection{The Performance of Metrics}
To evaluate the performance of the metrics, the Pearson linear correlation coefficient (PLCC)~\cite{PLCCRMSE}, the Spearman Rank Order Correlation Coefficient (SRCC)~\cite{SRCC} and the Root Mean Square Error (RMSE)~\cite{PLCCRMSE} are calculated. A five-parameter logistic fitting function proposed by the video quality experts group~\cite{5parameter} is used to map the dynamic range of scores from the objective metrics to a common scale, in order to ensure consistency between the various objective metrics and MOS. The performance of the metrics is shown in Table~\ref{tab:Metrics}.

\begin{table}[htbp]
\vspace{-1em}
\caption{The Performance of Metrics}
\begin{center}
\resizebox{\linewidth}{!}{
\begin{tabular}{|c|c|c|c|c|c|}
\hline
\textbf{Metrics} & \textbf{Reference} & \textbf{SRCC} & \textbf{PLCC} & \textbf{RMSE} \\
\hline
PSNR & $\checkmark$  & 0.75 & 0.76 & 1.70 \\
\hline
SSIM & $\checkmark$  & 0.77 & 0.83 & 1.46 \\
\hline
LPIPS (VGG) & $\checkmark$  & 0.74 & 0.75 & 1.72 \\
\hline
LPIPS (Alex) & $\checkmark$  & 0.77 & 0.81 & 1.51 \\
\hline
MSSSIM & $\checkmark$  & 0.76 & 0.83 & 1.43 \\
\hline
IW-SSIM & $\checkmark$  & 0.82 & 0.91 & 1.08 \\
\hline
VIF & $\checkmark$  & 0.79 & 0.85 & 1.48 \\
\hline
FSIM & $\checkmark$  & 0.76 & 0.79 & 1.59 \\
\hline
DISTS & $\checkmark$  & 0.86 & 0.88 & 1.24 \\
\hline
BRISQUE & $\times$  & 0.40 & 0.38 & 2.38 \\
\hline
CLIP-IQA & $\times$  & 0.64 & 0.61 & 2.08 \\
\hline
NIQE & $\times$  & 0.61 & 0.65 & 1.99 \\
\hline
FovVideoVDP & $\checkmark$ & 0.77 & 0.79 & 1.60 \\
\hline
VQM         & $\checkmark$ & 0.70 & 0.73 & 1.79 \\
\hline
VMAF        & $\checkmark$ & 0.83 & 0.83 & 1.44 \\
\hline
\end{tabular}}
\label{tab:Metrics}
\end{center}
\vspace{-2em}
\end{table}

It can be observed that the best full-reference metric is IW-SSIM with PLCC, SRCC and RMSE are 0.82, 0.91 and 1.1. We assume that its information-weighting mechanism is well suited for evaluating NeRF rendering considering that it can distinguish and emphasize distortions that occur in information-rich areas, such as fogging and surface scattering distortion. DISTS has higher PLCC than IW-SSIM, e.g., 0.86, but lower SRCC (0.88). All the non-reference metrics report obviously poor performance than other metrics. VMAF is the best video metric, it also shows good performance and wide acceptance on mesh and point cloud PVSs \cite{yang2023tsmd}. Using video metrics like VMAF is a good choice since we find that many types of artifact can only be well observed when displayed via a video form, such as wave, plane truncation, and floating distortion.


\section{Conclusion}
In this work, we create a new dataset called Explicit-NeRF-QA. It consists of 22 3D objects with diverse content and varying levels of compression distortion based on 4 explicit NeRF methods. This dataset undergoes comprehensive analysis to validate its sample diversity, MOS accuracy, and also illustrates typical NeRF distortion characteristics.  Based on the evaluation of SOTA metrics, the best full-reference metrics achieve a correlation around 0.85, while all the non-reference metrics are struggle with NeRF quality prediction with only 0.4 to 0.6 correlations. With accurate and large-scale MOS labels, this dataset enables the design of NeRF objective metrics and facilitates the development of NeRF compression technologies.

\bibliographystyle{IEEEtran}
\bibliography{IEEEabrv,ref}

\end{document}